\documentclass{article}

\usepackage{arxiv}

\usepackage[utf8]{inputenc} 
\usepackage[T1]{fontenc}    
\usepackage{hyperref}       
\usepackage{url}            
\usepackage{booktabs}       
\usepackage{amsfonts}       
\usepackage{nicefrac}       
\usepackage{microtype}      
\usepackage{lipsum}		
\usepackage{graphicx}
\usepackage{siunitx}
\usepackage{natbib}
\usepackage{amsmath}
\usepackage{authblk}

\newcommand{\unstable}{\mathbf{x}^{\text{unstable}}}
\newcommand{\sigreg}{\sigma_{\mathrm{reg}}}

\title{Interpreting and Stabilizing Machine-learning Parametrizations of Convection}

\author[1]{Noah D. Brenowitz}
\author[2,3]{Tom Beucler}
\author[2]{Michael Pritchard}
\author[1]{Christopher S. Bretherton}

\affil[1]{Vulcan, Inc, Seattle, WA\authorcr
  \{\tt noahb, chrisbr\}@vulcan.com}
\affil[2]{Department of Earth System Science, University of California, Irvine, CA\authorcr
\{\tt tbeucler, mspritch\}@uci.edu}
\affil[3]{Department of Earth and Environmental Engineering, Columbia University, New York, NY}

\begin{document}
\maketitle

\begin{abstract}
Neural networks are a promising technique for parameterizing sub-grid-scale physics (e.g. moist atmospheric convection) in coarse-resolution climate models, but their lack of interpretability and reliability prevents widespread adoption. For instance, it is not fully understood why neural network parameterizations often cause dramatic instability when coupled to atmospheric fluid dynamics. This paper introduces tools for interpreting their behavior that are customized to the parameterization task. First, we assess the nonlinear sensitivity of a neural network to lower-tropospheric stability and the mid-tropospheric moisture, two widely-studied controls of moist convection. Second, we couple the linearized response functions of these neural networks to simplified gravity-wave dynamics, and analytically diagnose the corresponding phase speeds, growth rates, wavelengths, and spatial structures. To demonstrate their versatility, these techniques are tested on two sets of neural networks, one trained with a super-parametrized version of the Community Atmosphere Model (SPCAM) and the second with a near-global cloud-resolving model (GCRM). Even though the SPCAM simulation has a warmer climate than the cloud-resolving model, both neural networks predict stronger heating/drying in moist and unstable environments, which is consistent with observations. Moreover, the spectral analysis can predict that instability occurs when GCMs are coupled to networks that support gravity waves that are unstable and have phase speeds larger than \SI{5}{\m\per\s}. In contrast, standing unstable modes do not cause catastrophic instability. Using these tools, differences between the SPCAM- vs. GCRM- trained neural networks are analyzed, and strategies to incrementally improve both of their coupled online performance unveiled.
\end{abstract}


\section{Introduction}

Global climate models (GCMs) still cannot both explicitly resolve convective-scale motions and perform decadal or longer simulations \citep{IPCC}. To permit grid spacings of \SI{25}{km} or larger, important physical processes operating at smaller spatial scales, such as moist atmospheric convection, must be approximated. This task is known as sub-grid-scale parameterization, and is one of the largest sources of uncertainty in estimating the future magnitude and spatial distribution of climate change \citep{Schneider2017-js}.

Owing to advances in both computing and available datasets, machine learning (ML) is now a viable alternative for traditional parameterization.
Viewed from the perspective of ML, parameterization is a straightforward regression problem.
A parameterization maps a set of inputs, namely atmospheric profiles of humidity and temperature, to some outputs, profiles of sub-grid heating and moistening.
\citet{Krasnopolsky2005-ca} and \citet{Chevallier1998-su} pioneered this growing sub-field by training emulators of atmospheric radiation parameterizations.
\citet{OGorman2018-hn} trained a random forest (RF) to emulate the convection scheme of an atmospheric GCM and were able to reproduce its equilibrium climate.
More recently, neural networks (NNs) have been trained to predict the total heating and moistening of more realistic datasets including the super-parametrized community atmosphere model (SPCAM) \citep{Rasp2018-ff,Gentine2018-kn} and a near-global cloud-resolving model (GCRM) \citep{Brenowitz2018-td,Brenowitz2019-qs}.

RFs appear robust to coupling: their output spaces are bounded since their predictions for any given input are averages over actual samples in the training data \citep{OGorman2018-hn}. In contrast, NNs are often numerically unstable when coupled to atmospheric fluid mechanics. In the case of coarse-grained GCRM data, \citet{Brenowitz2019-qs} eliminated an instability by ablating (i.e. removing) the upper-atmospheric humidity and temperature, which are slaved to the convective processes below, from the input space. \citet{Rasp2018-ff} also encountered instability problems, which they solved by using deeper architectures and intensive hyperparameter tuning, but instabilities returned when they quadrupled the number of embedded cloud-resolving columns within each coarse-grid cell of SPCAM, albeit without substantial retuning of the NN.

These sensitivities suggest that numerical instability can be related to ambiguity in the input data or imperfect choices of network architecture and hyperparameters.
Consistent with the former view, \citet{Brenowitz2018-td} argue that a NN may detect a strong correlation between upper-atmospheric humidity and precipitation, which is used by the parameterization in a causal way (humidity affects precipitation) when the true causality is likely reversed. On the other hand, the instabilities in SPCAM do not appear to be sensitive to this causal ambiguity and are not yet fully understood, but sensitivities to hyperparameter tuning are suggestive. Regardless of its origin, for NNs, the numerical stability problem is catastrophic because current architectures can predict unbounded heating and moistening rates once they depart the envelope of the training data, motivating our first question: \textit{can we unambiguously predict the stability of NN parameterizations of convection before coupling them to GCMs?}

Predicting the behavior of NNs is tied to the difficult problem of interpreting NN emulators of physical processes. While many interpretability techniques can be applied to NNs, such as permutation importance or layer-wise relevance propagation \citep[e.g.,][]{McGovern2019-op,toms2019physically,Montavon2018-dn,samek2017explainable,molnar2018iml}, we need to adapt these techniques to interpret NN parameterizations of convection. This motivates our second question: \textit{How can we tailor ML interpretability techniques, such as partial-dependence plots and saliency maps, for the particular purpose of interpreting NN parameterizations of convection?}

In atmospheric sciences, a common way to analyze convective dynamics utilizes the linearized response of parametrized or explicitly-resolved convection to perturbations from equilibrium \citep{Beucler2018-cd,Kuang2018-wh,Kuang2007-ph,Herman2013-iv}. These linearized response functions (LRFs) are typically computed by perturbing inputs in some basis and reading the outputs \citep[appendix B of][]{Beucler2019-lv} or by perturbing the forcing and inverting the corresponding operator \citep{Kuang2007-ph}. If the input/output bases are of finite dimension, then the LRF can be represented by a matrix. LRFs can also be directly computed from data, for instance by fitting a linear regression model between humidity/temperature and heating/moistening, or likewise by automatic differentiation of nonlinear regression models \citep{Brenowitz2019-qs}.

Visualizing the LRF as a matrix does not predict the consequences of coupling the scheme to atmospheric fluid mechanics (i.e., the GCM's "dynamical core"). \citet{Kuang2018-wh} takes this additional step, by coupling CRM-derived LRFs with linearized gravity wave dynamics.  He discovers convectively-coupled wave modes that differ from the linearly unstable eigen-modes of the LRFs. This 2D linearized framework has long been used to study the instability of tropical waves \citep{Hayashi1971-pc,Majda2001-gq,Khouider2006-ty,Kuang2008-jp}, but typically by analyzing a vertically-truncated set of equations for 2--3 vertical modes. Coupling the full LRF calculates these waves modes for the fully-resolved basis of vertical structures and allows the analysis to be easily performed for different base-states without recalculating the basis functions.

While LRFs provide a complete perspective on the sensitivity of a given parameterization, they can still be difficult to interpret because they have high dimensional input and output spaces.
Each side of the LRF matrix is equal to the number of vertical levels times the number of variables. However, the dominant process parametrized by ML schemes is moist atmospheric convection, which has well-known sensitivities to two environmental variables: the mid-tropospheric moisture and the lower-tropospheric stability (LTS). On the one hand, the intensity of convection increases exponentially with the former \citep{Bretherton2004-jf,Rushley2018-me}, perhaps because the buoyancy of an entraining plume is strongly controlled by the environmental moisture \citep{Ahmed2018-yd}. On the other hand, convection will fail to penetrate stable air, so LTS helps control the height of convection. 
Indeed, sufficiently large LTS is a prerequisite for forming stratocumulus cloud layers \citep{Wood2006-ro}. While the motions in shallow turbulently-driven clouds are not resolved at the \SI{1}{km} to \SI{4}{km} resolution of SPCAM or GCRM training datasets, we still expect lower stability to increase the depth of convection.

In this study, we probe the behavior of the NN parameterizations from our past work using these interpretability techniques. The main goals of this study are to 1) build confidence that ML parameterizations behave like realistic moist convection and 2) introduce a diagnostic framework that can predict if a NN will cause numerical instability. We will subject two sets of NN parameterizations to this scrutiny. The first set was trained by coarse-graining a GCRM simulation \citep{Brenowitz2019-qs} while the second set was trained using SPCAM \citep{Rasp2018-ff}. We will use the interpretable ML toolkit to compare the sensitivities of these two schemes and by extension the SPCAM and GCRM models.

The outline of the paper follows. 
In Section \ref{sec:ml-param}, we introduce the GCRM and SPCAM training datasets, and briefly summarize our corresponding ML parameterizations, which are quite similar in form. Then, Section \ref{sec:interpret} introduces the LTS-moisture sensitivity framework (Section \ref{sec:vary-lts-q}) and the wave-coupling methodology (Section \ref{sec:lrf+wave+intro}). The latter is used to assess the stability of NN parameterizations, so we need to describe the techniques we use to stabilize such schemes (Section \ref{sec:regularization}). Then, we present the results in Section  \ref{sec:results}. Section \ref{sec:lts-q-results} compares how changes in LTS and moisture control the NN parameterizations, while Sections \ref{sec:stabilizing-an-ml} and \ref{sec:res-stable=sp} apply the wave-coupling framework to predict coupled instabilities. We conclude in Section \ref{sec:conclusion}.

\section{Machine Learning Parameterizations}
\label{sec:ml-param}
\subsection{Global cloud-resolving Model\label{sec:GCRM}}

\citet{Brenowitz2018-td,Brenowitz2019-qs} trained their NNs with a near-global aquaplanet simulation performed with the System for Atmospheric Modeling (SAM) version 6.10 \citep{Khairoutdinov2003-du}.
This simulation is run in a tropical channel (from 23S to 23N) with a horizontal grid spacing of 4 km and 34 vertical levels of varying thickness, over a zonally symmetric ocean surface with a sea surface temperature of 300.15 K at the equator and 278.15 K at the poleward boundaries.  This GCRM training data consists of 80 days of instantaneous three-dimensional fields from this simulation, sampled every three hours.

The NN scheme parametrizes the apparent heating and moistening over 160 km grid boxes, a 40-fold downsampling of the original 4 km data.
This resolution is large enough that the precipitation still has significant autocorrelation at a lag of 3 hours, so the data are sampled at a high enough frequency to resolve some of the relevant moist dynamics.
The apparent heating $Q_1$ and moistening $Q_2$ are defined in terms of SAM's prognostic variables: the total non-precipitating water mixing ratio $q_T$ (kg/kg) and the liquid-ice static energy $s_L$ (J/kg).
On the coarse grid, these variables are advected by the large-scale flow and forced by the apparent heating $Q_1$ (W/kg) and moistening $Q_2$ (kg/kg/s).
These dynamics are described by:
\newcommand{\gcmten}[1]{\left( \frac{\partial\overline{#1}}{\partial t} \right)_{\text{GCM}}}
\begin{align}
  \frac{\partial \overline{s_L}}{\partial t} &= \gcmten{s_L} + Q_1,\label{eq:q1} \\ 
  \frac{\partial \overline{q_T}}{\partial t} &= \gcmten{q_T} + Q_2, \label{eq:q2}
\end{align}
where $\overline{\cdot}$ denotes a coarse grid average.

Unlike with SPCAM (see below), the apparent sources for the GCRM are defined as budget residuals by estimating the terms in (\ref{eq:q1}) and (\ref{eq:q2}).
The storage terms on the left hand side are estimated using a finite difference rule in time with the 3-hourly data.
The tendencies due to the coarse-resolution GCM are given by $\left(\partial\overline{s_{L}}/\partial t\right)_{\mathrm{GCM}}
 $  and $\left(\partial\overline{q_{T}}/\partial t\right)_{\mathrm{GCM}} $. 
They are estimated by initializing our `GCM', the coarse-resolution SAM (cSAM) at a grid spacing of \SI{160}{km} with the coarse-grained data, running it forward ten \SI{120}{s} time steps without any parametrized physics, and computing the time derivative by finite differences.
cSAM is run with a resolution of \SI{160}{km} as the coarse-graining time-scale.
For more details about this complex workflow, we refer the interested reader to \citet{Brenowitz2019-qs}.


\subsection{Super-parametrized Climate Model}

To complement the NN parameterization trained on the SAM global cloud-resolving
model, we analyze NNs trained on the Super-parametrized Community Atmosphere
Model v3.0 (SPCAM) \citep{Khairoutdinov2001-et,Khairoutdinov2005-zy}. 
SPCAM embeds eight columns of SAM (spatiotemporal resolution of
$4\;\textnormal{km}\times20\textnormal{s} $) in each grid column of the Community
Atmosphere Model (CAM, spatiotemporal resolution of
$2^{\circ}\times30\textnormal{min} $) in place of its usual deep convection and
boundary layer parameterizations to improve the representation of convection,
turbulence and microphysics.  In essence, SPCAM is a compromise between the
numerically-expensive global SAM and the overly-coarse CAM, which struggles to
represent convection \citep[e.g.,][]{oueslati2015double}. 
The goal of the NN parameterization \citep{Rasp2018-ff,Gentine2018-kn} is to emulate how the embedded SAM models vertically redistribute
temperature (approximately $Q_1$) and water vapor (approximately $Q_2$) in response to given coarse-grained conditions from
the host model's primitive equation dynamical predictions (i.e., temperature
profile, water vapor profile, meridional velocity profile, surface pressure,
insolation, surface sensible heat flux, and surface latent heat flux; all
prior to convective adjustment). The NN also includes the effects of cloud-radiative feedback by predicting the total diabatic tendency (convective plus radiative). 
Notable differences from the NN parameterization of SAM (section \ref{sec:GCRM})
include training on higher frequency data (30-minute instead of 3-hourly) of a
different form, in which a unambiguous separation between grid-scale drivers
and subgrid-scale responses can be exploited. 
This facilitates the NN parameterization's definition by avoiding the challenges
of coarse-graining. 
The lower computational cost of SPCAM, through its strategic undersampling of
horizontal space, also allows a longer duration training dataset (2 years
instead of 100 days for SAM) and the spherical dynamics of its host
model permit fully global (including extratropical) effects. 
The main disadvantage of SPCAM-based NNs is that superparameterization by
definition draws an artificial scale separation that inhibits some modes of
dynamics, and the idealizations of its embedded CRMs (e.g. 
2D dynamics and limited extent of each embedded array) compromise aspects of the
emergent dynamics \citep{pritchard2014restricting}. 
We refer the curious reader to section 2 of \citet{rasp2019online} for an extensive comparison of various ML parameterizations of convection.

\subsection{Neural network Parameterization}

The parameterizations for both SPCAM and the GCRM take a similar form.
A neural network predicts $Q_1$ and $Q_2$ as functions of the thermodynamic state within the same atmospheric column.
The parameterizations therefore have the following functional form
\begin{equation}
    \mathbf{Q} = \mathbf f(\mathbf{x}, \mathbf{y}; \theta); \label{eq:param-function}
\end{equation}
where $\mathbf{Q}= [Q_1(z_1),\ldots,Q_1(z_n),Q_2(z_1),\ldots,Q_2(z_n)]^T$ is a vector of the heating and moistening for a given atmospheric column; $\mathbf{x}$ is similarly concatenated vector of the thermodynamic profiles---$q_T$ and $s_L$ in the case of GCRM, or humidity and temperature for SPCAM; $\mathbf{y}$ are auxiliary variables such as sea-surface temperature, the insolation at the top of atmosphere for the GCRM, or surface enthalpy fluxes for SPCAM.
The ML will not prognose the source of these auxiliary variables.

Both \citet{rasp2018deep} and \citet{Brenowitz2019-qs} represent $f$ as a simple multi-layer feed-forward NN.
The hyperparameters and structures of their respective networks differ slightly (e.g. number of layers, activation functions), but in this article, we will only rely on the fact that NNs are almost-everywhere differentiable, a key advantage of NNs over other techniques (e.g. tree-based models).
NNs are almost-everywhere differentiable because they compose several affine transformations with nonlinear activations in between.
One ``layer'' of such an NN transforms the hidden values $x^{n}$ at the $n^{th}$ layer into the next layer's ``activations'' using the following functional form;
\begin{equation}
    \mathbf x^{n+1} = \sigma(A^n \mathbf x^{n} + \mathbf b^n),
\end{equation}
where $A^n$ is ``weight'' matrix, $\mathbf{b}^n$ is a "bias" with the same size as $\mathbf{x}^n$, and $\sigma$ is a nonlinear activation function that is applied elementwise on its input vector.
The inputs feed into the first layer---$\mathbf x^0 = \mathbf{x}$---and the outputs read out from the final layer $\mathbf{x}^m=\mathbf{Q}$.
The parameters of the NN are the collection of weight matrices and bias vectors that we mathematically denote using a single vector $\theta =\{A_1, \ldots, A_m, b_1, \ldots, b_m\}$ where $m$ is the total number of layers.

The parameters $\theta$ are tuned by minimizing a cost function $J(\theta)$, typically a weighted mean-squared error, using stochastic gradient descent.
Modern NN libraries such as Tensorflow \citep{tensorflow2015-whitepaper} or PyTorch \citep{PyTorch} enable such a training procedure by automatically computing derivatives of functions like (\ref{eq:param-function}) with respect to their parameters.
In this paper, we will also use this capability to explore the linearized sensitivity of a NN parameterization to its inputs across a wide array of base states.

For the GCRM, we analyze a NN that initially includes inputs from all vertical levels of humidity and temperature, a configuration that causes a prognostic simulation to crash after 6 days \citep{Brenowitz2019-qs}.
The network has 3 hidden layers of 256 nodes each and uses ReLU activation, and is trained for 20 epochs using the Adam optimizer to minimize the mass-weighted mean-squared-error of predicting the $Q_1$ and $Q_2$ estimated by budget residuals of (\ref{eq:q1}) and (\ref{eq:q2}).

For SPCAM, we analyze two NNs with
identical architectures and training conditions (9 fully-connected layers of 256
nodes each trained for 20 epochs using the Adam optimizer): ``NN-stable'' and
``NN-unstable''. 
Although both NNs were trained using $\sim140\textnormal{M} $ samples from
aquaplanet simulations, the training simulation for NN-stable used 8 SAM columns
per grid cell and underwent significant hyperparameter tuning while the training
simulation for NN-unstable used 32 SAM columns per grid cell and the NN was less
intensively tuned.  Helpfully for our purposes, NN-stable led to successful multi-year
climate simulations once prognostically coupled to CAM \citep{Rasp2018-ff}, but NN-unstable proved prone to producing moist mid-latitude
instabilities that led all prognostic simulations to crash within 2-15 days (see Movie S1). While the time to crash was sensitive to initial condition details,
no simulations with NN-unstable proved capable of running more than 2 weeks.

In the following sections, we show how physically-motivated diagnostic tools
help anticipate, explain, and begin to resolve these problematic instabilities.

\section{Interpreting ML parameterizations \label{sec:interpret}}

\subsection{Variation of Inputs \label{sec:vary-lts-q}}

A few important parameters control the strength and height of moist
atmospheric convection. Any parameterization, including a machine
learning parameterization, should capture the dependence of convection
to these parameters. One such parameter is the lower tropospheric
stability
\[
LTS=\theta(700\;\mathrm{hPa})-SST
\]
where $\theta$ is the potential temperature and SST is the sea-surface temperature.
Low LTS indicates the lower troposphere is conditionally unstable, favoring deep convection. 

A second controlling parameter is the lower-tropospheric moisture, defined
by
\[
Q=\int_{850}^{550}q_{T}\frac{dp}{g}.
\]
Cumulus updrafts entrain surrounding air as they rise through the lower troposphere.  If that air is dry (low $Q$), the entrained air induces considerable evaporative cooling as it mixes into the cloudy updraft, impeding deep precipitating convection.  Hence
moist columns tend to precipitate exponentially more than dry ones
\citep{Bretherton2004-jf, Neelin2009-dg,Ahmed2018-yd,Rushley2018-me}.

To see how the NNs depend on these important control parameters, both the GCRM and SPCAM training datasets are partitioned into bins of LTS and $Q$.
For the GCRM training dataset, we partition the points in the tropics and subtropics (23S -- 23N); the humidity bins are \SI{2}{mm} wide, and the LTS bins are \SI{.5}{K} wide.
For SPCAM, we use 20 bins evenly spaced between \SI{0}{mm} and \SI{40}{mm} for mid-tropospheric moisture and from 7K to 23K for LTS.
In both cases, the NN's inputs $x$ are averaged over these bins. Denote this average by $E[\bar{\mathbf{x}} | Q, \text{LTS}]$.
A parameterization's
sensitivity to the variables $Q$ and $S$ is given by
\begin{equation}
f(Q,S)=f(E[\bar{\mathbf{x}} | Q, S];\theta).\label{eq:bin-sensitivity}
\end{equation}
Because $f$ is nonlinear, this is not equivalent to taking the average
of the NN's outputs over the bins. Rather, it tests
the nonlinear sensitivity to systematic changes in its inputs.
In the sections below, we will also plot the bin-averages of the ``true'' apparent heating $E[Q_1 | Q, \text{LTS}]$ and moistening $E[Q_2 | Q, \text{LTS}]$, which indicate the fraction the true variability across bins the NN is able to predict successfully. 

\subsection{Linear Response Functions\label{sec:lrf}}

The method above shows how a ML parameterization depends nonlinearly
on a few inputs, but it is difficult to extend to the full input space
of a parameterization. To do this, we instead use the LRF or saliency map.
The
LRFs in this study will be computed from the output of a neural
network (NN). By using a nonlinear continuous (and almost everywhere
differentiable) model such as a NN, we can compute the local linear
response of convection for a variety of base-states.

LRFs have already been employed to develop machine learning parameterizations. For instance \citet{Brenowitz2019-qs} computed LRFs to analyze what was causing their neural network parameterizations to produce unstable dynamics when coupled to a GCM.
For most of this analysis, we linearize the GCRM-trained NN about the tropical mean profile.
This state is not a radiative-convective equilibrium (RCE), so some positive modes of the linearized response function
likely represents decay to a true RCE state. Developing machine learning
parameterizations with a stable radiative-convective equilibrium is
a task for future research.

\subsection{Coupling to Two-dimensional Linear Dynamics\label{sec:lrf+wave+intro}}

While LRFs provide insights into how a parameterization affects a single atmospheric column in radiative convective equilibrium, they cannot alone predict the feedbacks that will occur when coupled to the environment.
This coupling is thought to play a critical role in the catastrophic instability because NNs that produce accurate single column model simulations \citep{Brenowitz2018-td} can still blow up when coupled to the three-dimensional wind field simulated in a GCM \citep{Brenowitz2019-qs}.
While we ultimately suspect that nonlinearity causes coupled simulations to catastrophically blow up once the NNs are forced to make predictions outside of the training set, we hypothesize that the initial movement towards the edge of the training manifold is inherently linear. 
In particular, we suspect it arises from interactions between the parameterization and small-scale gravity waves, which are the fastest modes present in large-scale atmospheric dynamics.
This interaction has been extensively studied in the literature \citep{Hayashi1971-pc,Majda2001-gq}, and is known to cause deleterious effects such as grid-scale storms when using traditional parameterizations based on a moisture convergence closures \citep{emanuel1994atmospheric}.
This suggests the problem can be understood from a two-dimensional perspective, so we also neglect the role of more complicated three-dimensional dynamics and rotation.

We now derive the linearized dynamics of a ML parameterization coupled to gravity waves.
Following the above discussion, we assume that the flow is two-dimensional (vertical and horizontal), anelastic, and hydrostatic. 
Further assuming that the mean winds are zero, the linearized anelastic equations in terms of humidity $q$, static energy $s$, and vertical velocity $w$ perturbations are written as
\begin{align}
q_{t}+\bar{q}_{z}w & =Q_{2}',\\
s_{t}+\bar{s}_{z}w & =Q_{1}',\\
w_{t} & =-\left(A^{-1}B\right)_{xx}-dw. \label{eq:w} 
\end{align}
The final equation is obtained by taking the divergence of the horizontal momentum equation and eliminating the pressure gradient term using hydrostatic balance and the anelastic nondivergence condition (see Appendix (a) for more details).

Here, $A$ is a continuous  elliptical vertical differential operator defined by
$Aw=\frac{\partial}{\partial z}\left(\frac{1}{\rho_{0}}\frac{\partial}{\partial z}(\rho_{0}w)\right)$. When endowed with rigid lid boundary conditions, $w(0)=w(H)=0$, this linear operator can be inverted to give $A^{-1}$. In practice, we discretize these continuous operators using finite differences so that these operations can be performed with matrix algebra.

Since we are focused on free-tropospheric dynamics, we have neglected the virtual effect of water vapor and approximated the buoyancy by $B = g s/\bar{s}$. 
The momentum damping rate is fixed at $d=1/(\SI{5}{d})$.
We discretize this equation using centered differences
(more details in Appendix (\ref{sec:discrete-elliptic}), and assume rigid lid
($w=0$) boundary conditions at the surface and top of atmosphere like \citet{Kuang2018-wh}.

\newcommand{\state}{%
\begin{pmatrix} \mathbf{q} \\ \mathbf{s} \\ \mathbf{w}%
\end{pmatrix}}
\newcommand{\diag}{\mbox{diag}}

The perturbation heating $Q_1'$ and moistening $Q_2'$ are a linear transformation
of the perturbed state which could be non-local in the vertical direction.
In particular, 
\[
Q_{1}'(z)=\frac{1}{H}\int_{0}^{H}\left[\frac{\partial Q_{1}(z)}{\partial q(z')}q(z')+\frac{\partial Q_{1}(z)}{\partial s(z')}s(z')\right]dz',
\]
and similarly for $Q_{2}'$. Upon discretizing this integral onto
a fixed height grid and concatenating the $s$ and $q$ fields into vectors
$\boldsymbol{s}=[s(z_{1}),\ldots,s(z_{n})]^T$, and similarly for $q$, $Q_1$, and $Q_2$,
this continuous formula can be written as 
\[
\begin{pmatrix}
\mathbf {Q}_1 \\
\mathbf {Q}_2
\end{pmatrix} = \begin{pmatrix}
M_{ss} & M_{sq} \\
M_{qs} & M_{qq}
\end{pmatrix} \begin{pmatrix}
\mathbf{s}\\
\mathbf{q}
\end{pmatrix}
\].
The subblocks of the matrix (e.g. $M_{qq}$) encode the linearized response function
for the fixed height grid $z_{1},\ldots,z_{n}$.
Then, assuming the solution is a plane wave in the horizontal direction with a wavenumber $k$, (\ref{eq:w}) can be encoded using the following matrix form
\begin{equation}
    \frac{\partial}{\partial t} \state = {\cal T}\state, \label{eq:update}
\end{equation}
where the linear response operator ${\cal T} $ for total water, dry static energy and vertical velocity is given by:
\begin{equation}
 {\cal T} = \begin{pmatrix} 
    M_{qq} & M_{qs} & \diag(\mathbf{\bar{q}}_z) \\
    M_{sq} & M_{ss} & \diag(\mathbf{\bar{s}}_z) \\
    0 & -g k^2  A^{-1} \diag(\mathbf{\bar{s}})^{-1}  & -d I \\
    \end{pmatrix}.
    \label{eq:matrix}
\end{equation}
Here, $I$ is the identity matrix, and $\diag$ creates a matrix with a diagonal given by its vector argument.

The spectrum of (\ref{eq:matrix}) can be computed numerically for each wavenumber $k$ to obtain a dispersion relationship.
Appendix (\ref{sec:discrete-elliptic}) derives the discretization we use for the elliptic operator $A$.
The real component of any eigenvalue $\lambda$  of ${\cal T} $ is the growth rate, and the phase speed can be recovered using the formula $c_p = -\Im{\lambda} / k$. The eigenvectors of ${\cal T} $ describe the vertical structure of the wave mode in terms of $s$, $q$, and $w$. Let $\mathbf{v}$ be such an eigenvector, then the wave's structure over a complete phase of oscillation can be conveniently plotted in real numbers by showing $\Re \{\mathbf{v} \exp{i \phi}\} $ for $0 \leq \phi < 2 \pi$. 
In the sections below, we will show the phase speed and growth rates for every single eigenmode over a range of wavenumbers.
Then, we can visualize the vertical structures of a few particularly interesting modes, such as unstable propagating modes or standing waves.
 
\section{Regularization\label{sec:regularization}}

As we have seen, NNs are often numerically unstable when coupled to atmospheric fluid dynamics, and much of our recent work has focused on solving this central challenge.  One reason in the case of training data from coarse-grained simulations may be causality issues.  In the GCRM, there is a strong correlation between an input variable---upper tropospheric total water---and an output variable---precipitation.  This correlation would be expected because deep convection lofts moisture high into the atmosphere and total water includes cloud water and ice. However, using it as a closure assumption would violate a physical causality argument that moist atmospheric convection is triggered by lower atmospheric thermodynamic properties.

\citet{Brenowitz2019-qs} found that reducing the potential for spurious causality by ablating both the upper atmospheric temperature and humidity from the input features of an NN parameterization results in a stable scheme.  It is ad-hoc, but works consistently. 
This was discovered using a LRF analysis (see Sec. \ref{sec:lrf}), which demonstrates that ML interpretability techniques have already significantly aided the development of ML parameterizations.

For SPCAM trained NNs, stability has also been a lingering challenge. 
Unfortunately, removing upper atmospheric inputs as prescribed by \citet{Brenowitz2019-qs} did not stabilize these NNs (see e.g. Movie S2). We speculate that the instabilities from SPCAM are linked to inevitable imperfections of NN fit, exacerbated by limited hyperparameter tuning. Nonetheless, using some of the same interpretability techniques described above, we have developed an ``input regularization'' technique for stabilizing SP-trained NNs.

Just as with the upper atmospheric ablation for the GCRM NNs, this technique was discovered using a LRF analysis.
We noticed that
directly calculating the LRF of SPCAM-trained NNs via automatic differentiation
results in noisy, hard-to-interpret LRFs (top line of Figure S2).
When coupled
to 2D dynamics, these LRFs produce unphysical stability diagrams, with unstable
modes with phase speeds greater than \SI{300}{\m\per\s} even for the ``NN-stable'' network (bottom line of Figure S2).

The LRF's noisiness indicates that the NN responds nonlinearly to
small changes in its inputs.
While this behavior could be a desirable aspect of the NN convective
parameterization \citep{Palmer2001-rg}, it prevents a clean interpretation of
the NN parameterization through automatic differentiation about an individual
basic state alone. 
On the other hand, the SAM-trained NNs have a much smoother response, a discrepancy that could be due to differences in the network architecture, training strategy, or the underlying training data. Understanding this discrepancy between two networks with the same function trained on datasets with obvious similarities is an important future challenge.

The ``NN-stable'' SPCAM network performs stably when interactively coupled to GCM dynamics, suggesting it is not overly sensitive to larger perturbations. 
This motivates feeding an ensemble $\left({\cal X}\right) $ of randomly perturbed inputs to the SPCAM NN
parameterization, which then outputs an ensemble $\left({\cal Y}\right) $ of
predictions that we may average to more cleanly understand the NN's behavior.

For concreteness, we now apply ``input regularization'' to our initial problem, i.e. the unstable behavior of ``NN-unstable,'' recalling that it stems from an imperfect fit to an especially difficult training regime when 32-column CRMs are used in the SPCAM training dataset. This requires 5 steps:
\begin{enumerate}
    \item We track the (longitude, latitude) coordinates of the instability (see
      Movie S1) back in time to identify the base state
      $\boldsymbol{x_{\mathrm{unstable}}} $ leading to the instability. For
      simplicity, we choose the earliest timestep for which the perturbation
      responsible for the crash produces a maximum in the convective moistening
      field ($Q_2$).
    \item We construct an input ensemble $ \{{\mathbf{x}^{(i)}}_{i=1,2,\ldots,n}\} $
      of $n$ members by perturbing the input $\boldsymbol{x_{\mathrm{unstable}}} $ producing the instability using a normal distribution $\cal{N} $ of mean 0 and standard deviation $\sigma_{\mathrm{reg}} $:  
    \[
     x^{(i)}_j = (1 + z^{(i)}_j)\unstable_j ,\quad z^{(i)}_j \sim \mathcal{N}( 0, \sigreg).
    \]
    We refer to $\sigma_{\mathrm{reg}} $ as ``regularization amplitude'' (in $\% $): the larger $\sigma_{\mathrm{reg}} $ is, the broader the ensemble of exact inputs $\boldsymbol{x} $ will be in the input ensemble.  
    \item We feed each member $\mathbf{x}^{(i)}$ of the input ensemble into the
      NN parameterization, producing an ensemble of outputs $\{\mathbf{y}^{(i)}\}_{i=1,\ldots,n}$.
    \item We calculate the LRF about the input ensemble by automatically differentiating each
      input-output pair before taking the ensemble-mean of the resulting LRFs:  
      \[
       \text{LRF}_{\text{reg}}=\frac{1}{n} \sum_{i=1}^n \frac{\partial y^{(i)}}{\partial x^{(i)}}
      \]
    \item We couple the ``regularized'' LRF to our two-dimensional wave coupler to calculate a stability diagram representative of the NN's behavior near the ``regularized'' set of inputs $ \{{\mathbf{x}^{(i)}}_{i=1,2,\ldots,n}\} $.
\end{enumerate}

While we do not advocate for such an approach as a general strategy to stabilize SPCAM-based NNs---adding significant spread to the input vector will guarantee taking the surrogate model outside of its training set where large biases are expected---we will exploit this interesting property to investigate the robustness of our interpretability framework by comparing offline predictions to online prognostic failure modes. That said, adding Gaussian noise to inputs is commonly used to regularize NNs during training.
 
\section{Results\label{sec:results}}

\subsection{The Onset of Deep Convection in ML parameterizations \label{sec:lts-q-results}}

\begin{figure}
\includegraphics[width=\textwidth]{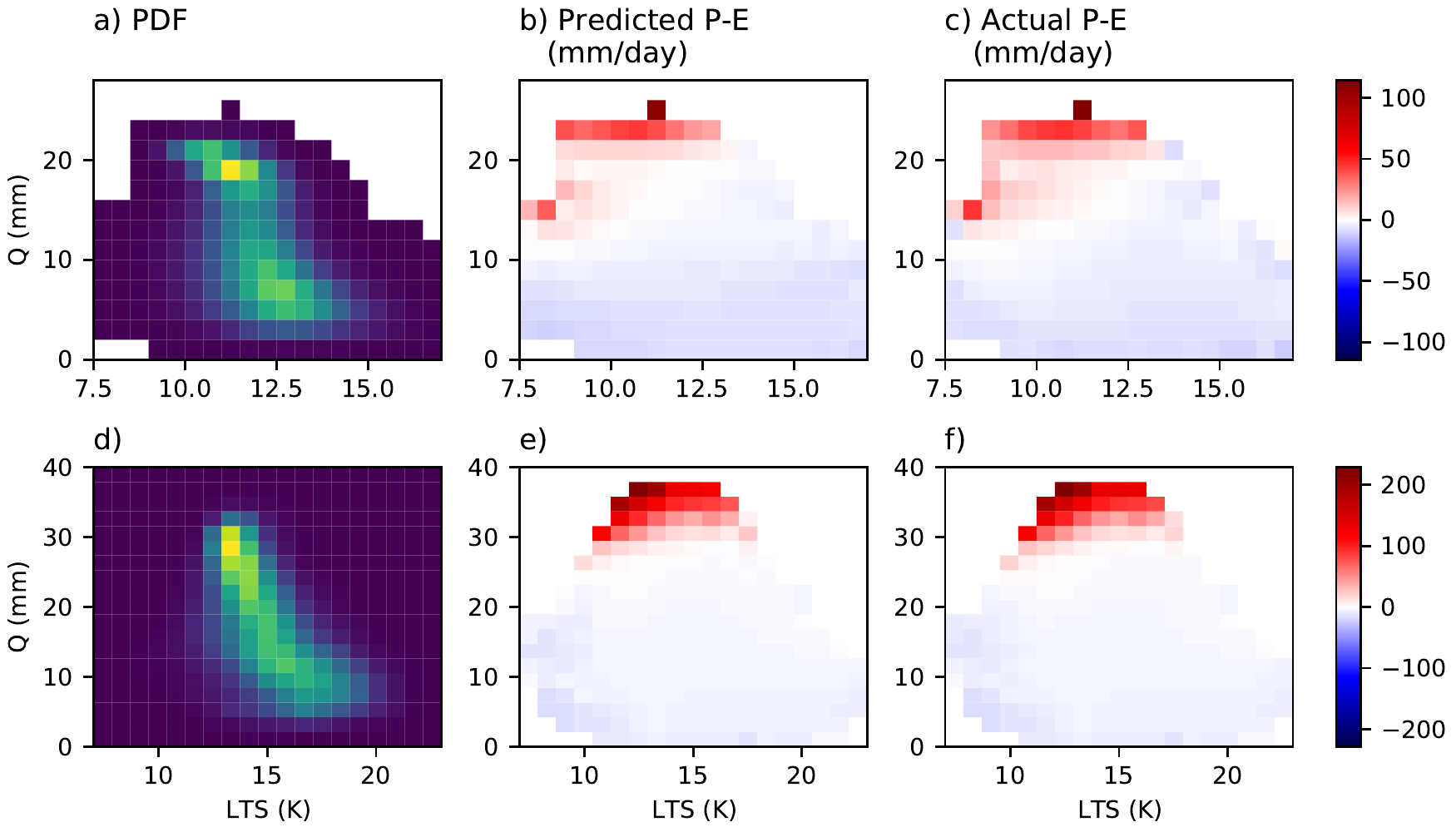}
\caption{Binning in LTS and Moisture Space. The first row shows, for the GCRM, (a) population of each LTS/Q bin, predicted net precipitation (b) and (c) the  $- \langle Q_2 \rangle \approx P - E$ from the training dataset. $\langle \cdot \rangle$ is the mass-weighted vertical integral. (d) - (f) are the corresponding results for SPCAM.\label{fig:lts-q-bins}}
\end{figure}

Figure \ref{fig:lts-q-bins} shows the two dimensional binning of the combined tropical and
subtropical data in mid-tropospheric humidity $Q$ and lower tropospheric
stability (LTS) space for all latitudes equatorward of 22.5 degrees.
For the GCRM simulation, the distribution is bi-modal, with many high-moisture low-stability samples---presumably from the tropics---and another peak for lower moistures of about 7 mm from the sub-tropics.
The SPCAM simulation is moister, with a modal $Q$ around \SI{30}{mm} compared to less than \SI{20}{mm} in the GCRM simulation.
This is not surprising because the peak SST of the SPCAM simulation is ~2-3 K warmer.

We use net precipitation (surface precipitation minus evaporation) as a proxy for deep convection, because it is predicted by both NNs as the column integral of the apparent drying $-Q_2$, and because it clearly distinguishes between regimes of little precipitation ($P < E$) and substantial precipitation ($P > E$).  Both training datasets, and the ML parameterizations trained from them, predict that the net precipitation increases dramatically with mid-tropospheric humidity. A similarly nonlinear dependence has been documented in numerous
studies \citep{Bretherton2004-jf,Neelin2009-dg,Rushley2018-me}. The net precipitation depends less strongly on LTS, but for intermediate values of moisture around 15-20 mm, the stability is an important control. The difference between the bin-averaged net precipitation and the NNs predictions with the bin-averages are relatively small ($\sim\SI{10}{mm/day}$). 
We conclude that the machine learning parameterizations depend smoothly and realistically on $Q$ and LTS.

\begin{figure}
\includegraphics[width=1.0\textwidth]{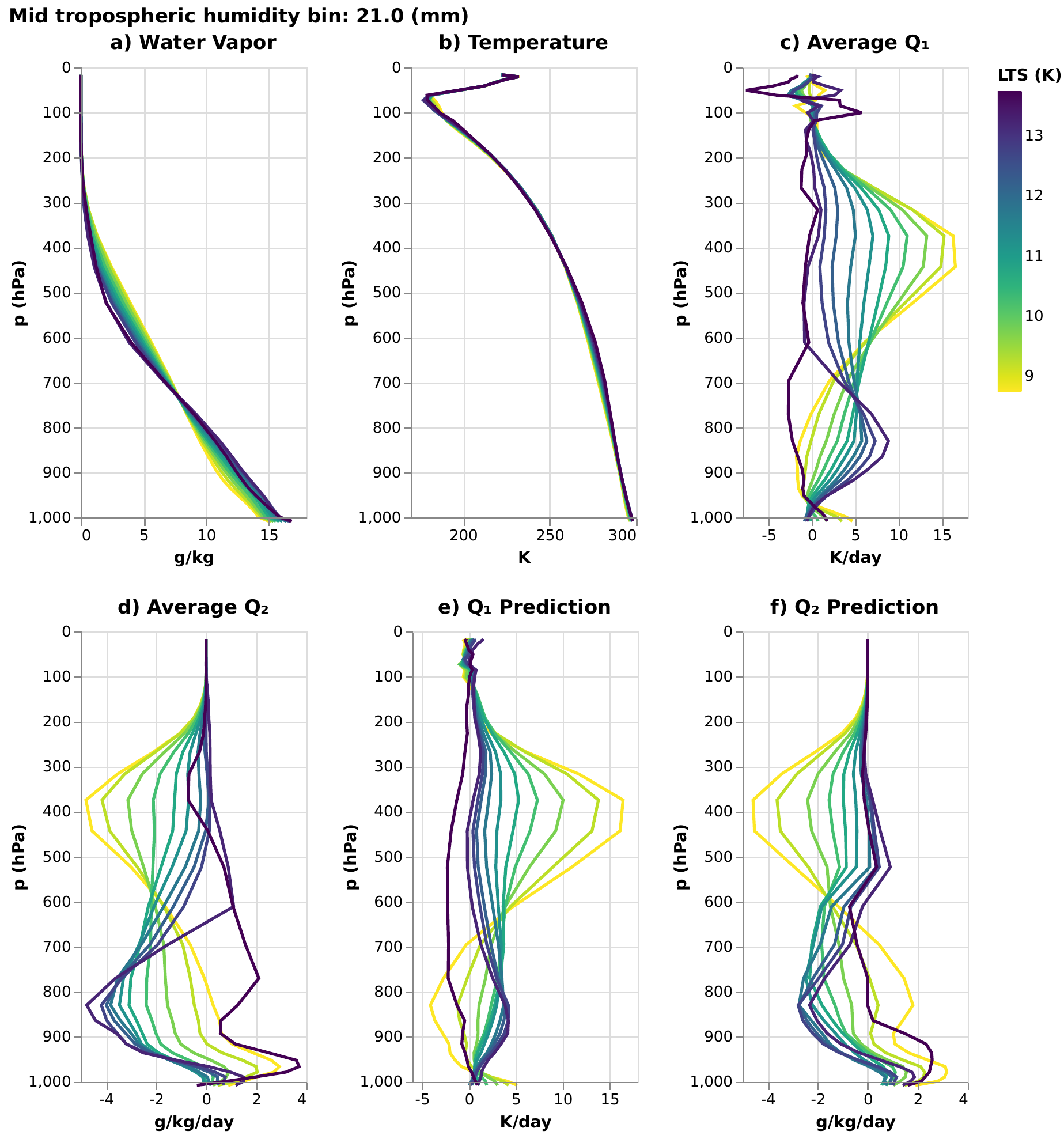}
\caption{Deepening of convection for SAM, varying the LTS with mid-tropospheric humidity
between 20 and 22 mm. \label{fig:vary-lts}}
\end{figure}

\begin{figure}
\centering
\includegraphics[width=1.0\textwidth]{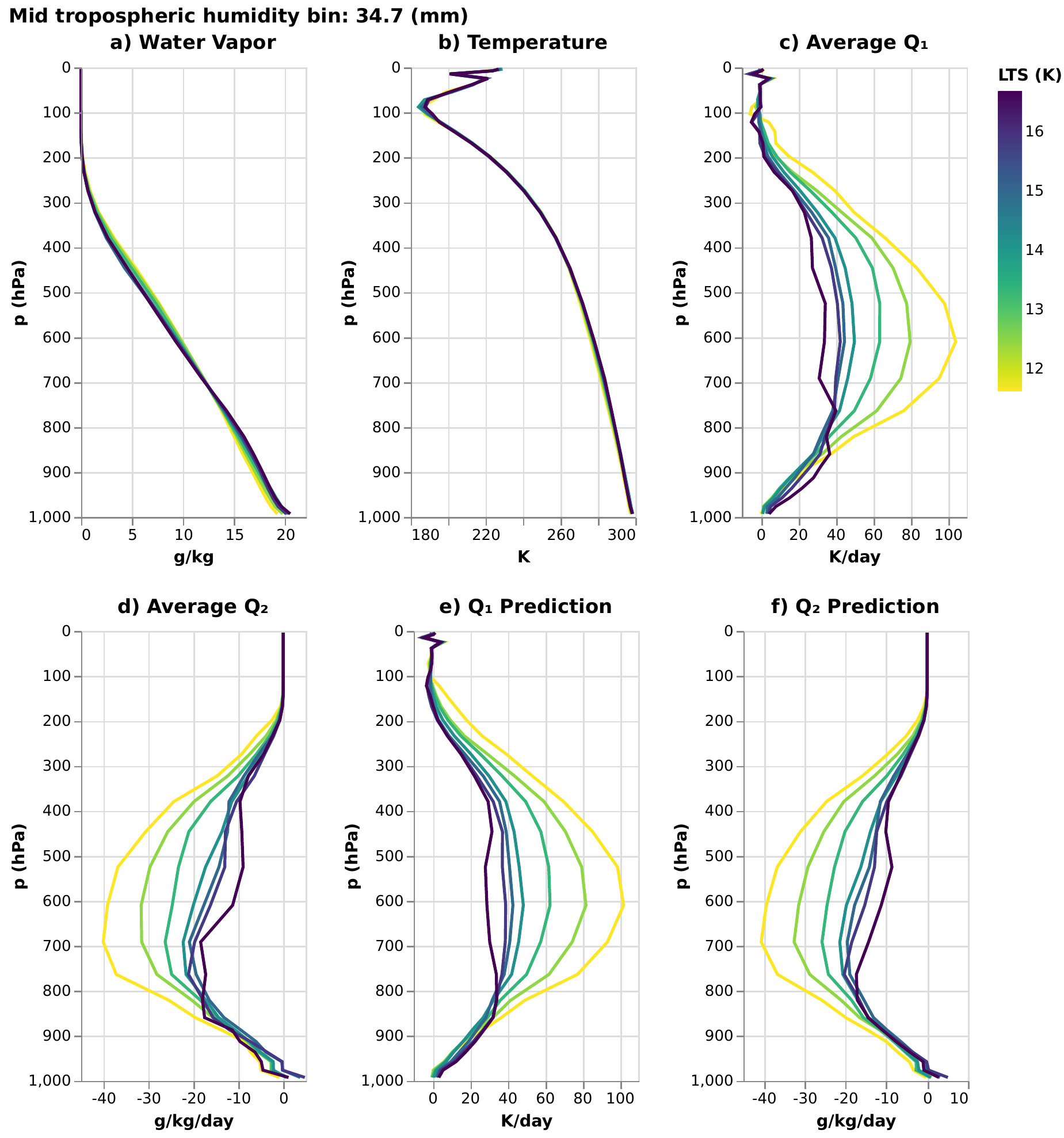}

        \caption{Deepening of convection for SPCAM, varying the LTS with mid-tropospheric humidity
between 33.7 and 35.7 mm. \label{fig:tomvarylts}}
\end{figure}

How does the vertical structure of the input variables and the predicted heating and moistening vary with $Q$ and LTS?
Figure \ref{fig:vary-lts} shows vertical profiles of these quantities for varying LTS binned over tropical grid columns with $20 \leq Q < \SI{22}{mm}$ for the GCRM simulation while Figure \ref{fig:tomvarylts} shows the LTS dependence for $33.7 \leq Q < \SI{35.7}{mm}$ for the moister SPCAM simulation. The humidity reaches much higher in the atmosphere for low stabilities, and the lower tropospheric must offset these gains to maintain a constant $Q$.
For the GCRM, the predicted heating and moistening switch from shallow to deep profiles as LTS decreases.
However, the overall magnitude of moistening is relatively unchanged, 
consistent with Figure \ref{fig:lts-q-bins}c. T
hus, LTS controls the height of the predicted convection more than its
overall strength.
That said, it is unclear whether these changes are a direct response to the lower-tropospheric temperature structure or controlled by the simultaneous changes in the humidity profile (Figures \ref{fig:vary-lts}a and \ref{fig:tomvarylts}a).
On the other hand, the SPCAM NNs fail to predict such a clear deepening of convection with decreased stability.
This could owe to the $Q$-bin we selected for this analysis or to a more fundamental differences in moist convection between the SPCAM and GCRM simulations.
Nonetheless, each NN faithfully represents the convective sensitivity of its own training dataset.

The predicted heating and moistening vary in a similar way to the bin-averaged $Q_1$ and $Q_2$ profiles, but the latter are more sensitive to the LTS than the NN.
Note that the NNs make their predictions with a single input profile whereas the bin-averaged $Q_1$ and  $Q_2$ are statistical averages of many individual heating and moistening profiles.
Thus, these figures demonstrate how the NN's prediction are sensitive to systematic changes in the input variables, whereas the bin-averaged heating and moistening show a statistical, but potentially non-causal link between heating, moistening, mid-tropospheric humidity, and LTS in the data.

\begin{figure}
\includegraphics[width=1.0\textwidth]{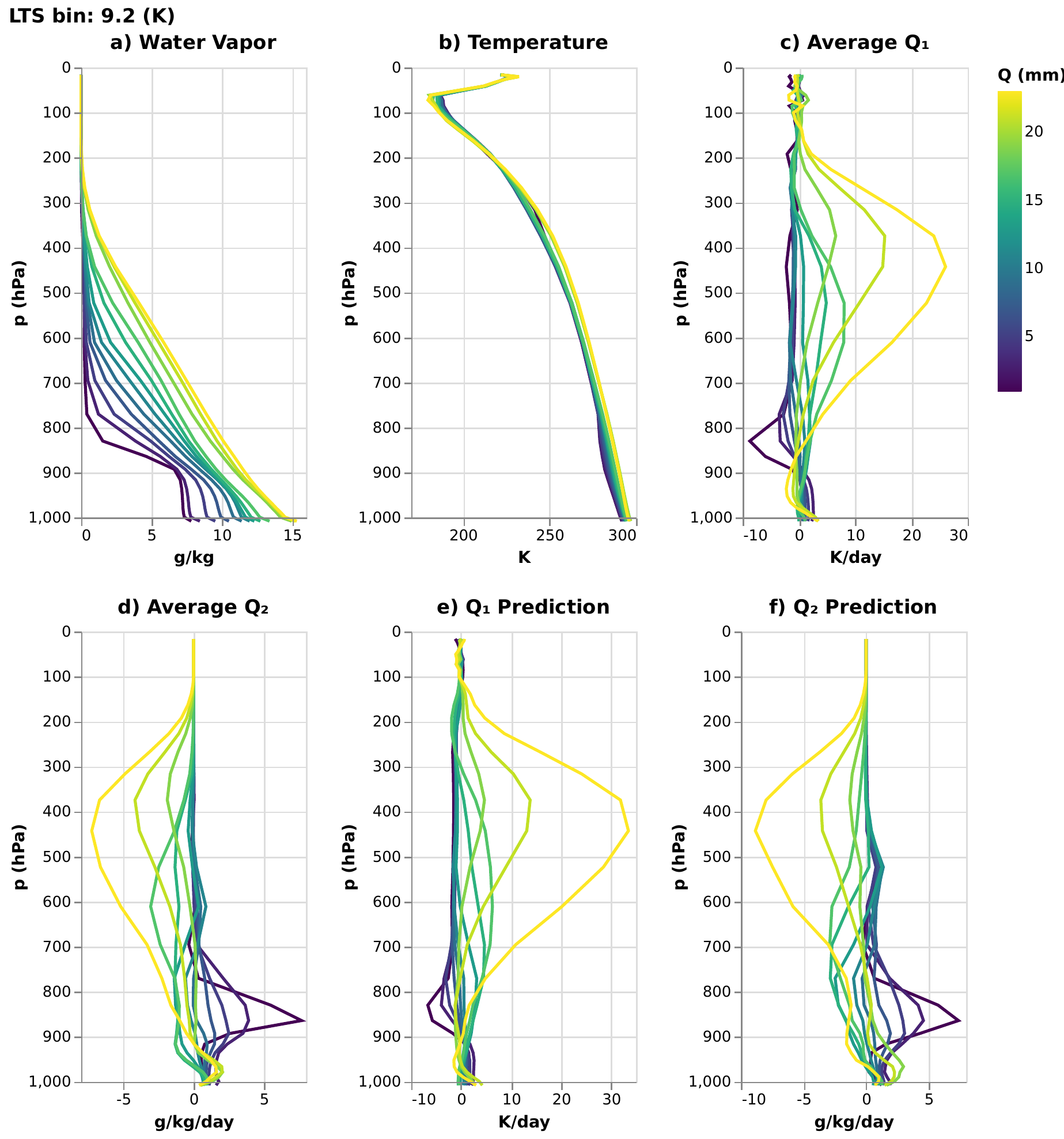}

\caption{Strengthening of convection for SAM, varying mid-tropospheric moisture
$Q$ for LTS between 9.0 and 9.5 K. \label{fig:vary-q}}
\end{figure}

\begin{figure}
    \centering
    \includegraphics[width=1.0\textwidth]{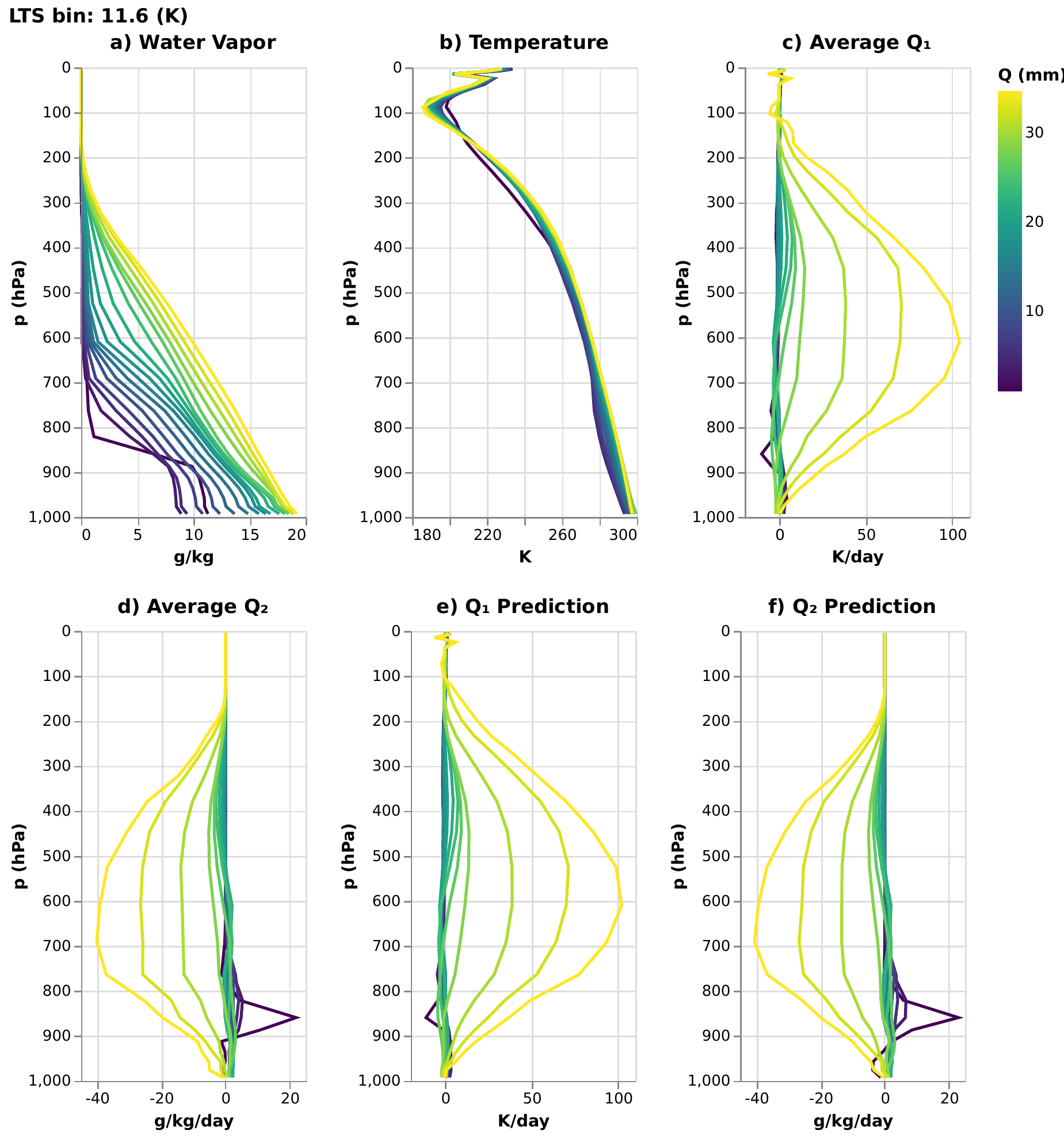}

    \caption{Strengthening of convection for SPCAM, varying mid-tropospheric moisture
$Q$ for LTS between 11 and 12 K. }
    \label{fig:tomvaryq}
\end{figure}

Figures \ref{fig:vary-q} and \ref{fig:tomvaryq} show how the mid-tropospheric humidity controls the
vertical structure of the input and response variables, for the GCRM and SPCAM data respectively. 
The LTS is fixed between \SIlist{9;10}{K} for the GCRM run and \SIlist{11;12}{K} for SPCAM, both of which are unstable ranges.
The overall amount of water changes dramatically for this change in both simulations because $Q$ is highly correlated with the total precipitable water in an atmospheric column.
Both NNs predict cooling and moistening near the top of the boundary layer for the drier profiles ($z=\SI{2000}{\km}$) due to shallow clouds.
Once a threshold $Q$ is reached, the sign flips and the machine learning
parameterization predicts increasingly deep heating and moistening.  For the three moistest profiles, the heating and moistening dramatically 
strengthen with little change in vertical structure. 
The predicting tendencies are again similar to their bin averages.

\subsection{Stabilizing via Ablation of Inputs}
\label{sec:stabilizing-an-ml}

\citet{Brenowitz2019-qs} obtained a stable scheme by training their NN without
input from the upper atmospheric temperature or humidity.
However, they did not examine whether it was ablating the atmospheric temperature, humidity or both
that prevented numerical instability.
The term ``ablate'' is used as an analogy to neuroscience research on how removing (i.e. ablating) brain tissue affects animal behavior.
In this section, we use the wave-coupling framework introduced in Section \ref{sec:lrf+wave+intro} to explore the independent effects of ablating temperature and humidity. 
Because this wave coupling is performed for one wavenumber at a time it is much more computationally affordable than a full non-linear simulation, but still hints at how the NN will perform in coupled simulations.

To study this further, we first compute the LRF of an NN trained with all
atmospheric inputs \cite[cf. Figure 1]{Brenowitz2019-qs}. 
Then, the upper atmospheric humidity and/or temperature inputs are sequentially
knocked out by inserting 0 in the corresponding entries of the LRF. 
This section studies the following configurations:
\begin{itemize}
\item All atmospheric inputs (unablated),
\item No humidity above level 15, 
\item No temperature above level 19,
\item No humidity above level 15 nor temperature above level 19 (fully-ablated).
\end{itemize}

\begin{figure}
\includegraphics{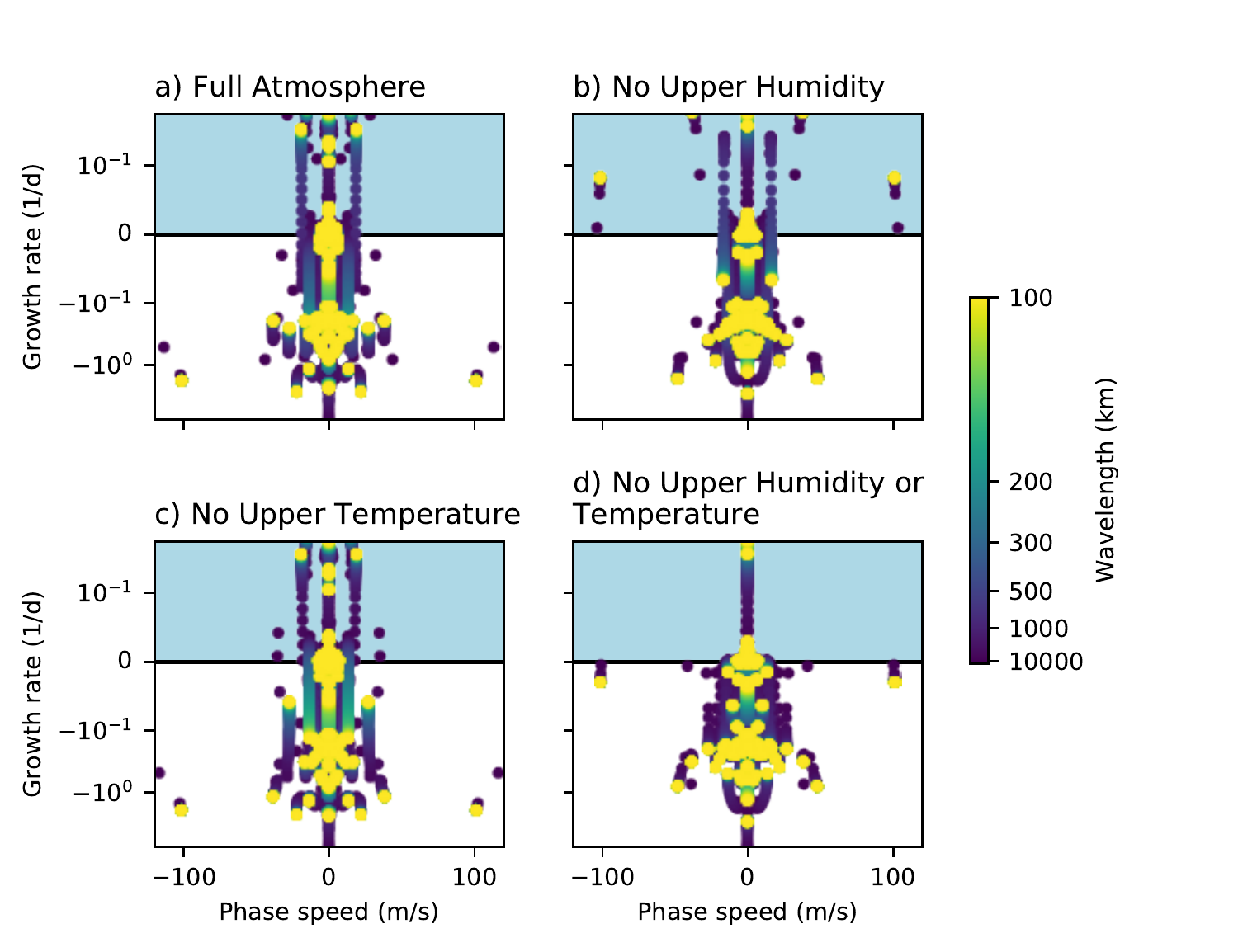}
\caption{Wave spectra with and without lower tropospheric input. a) full atmospheric input, b) lower-tropospheric humidity, c) lower-tropospheric temperature, and d) lower-tropospheric humidity and temperature. The light-blue background indicates where the phase speed is greater than $|c_p| > \SI{5}{m/s}$ and the growth rate is positive. This box is more visible in the zoomed-in plot (cf. Figure \ref{fig:spectra-zoom}). Waves inside of this region are likely responsible for instability. \label{fig:spectra}}
\end{figure}

\begin{figure}
\includegraphics{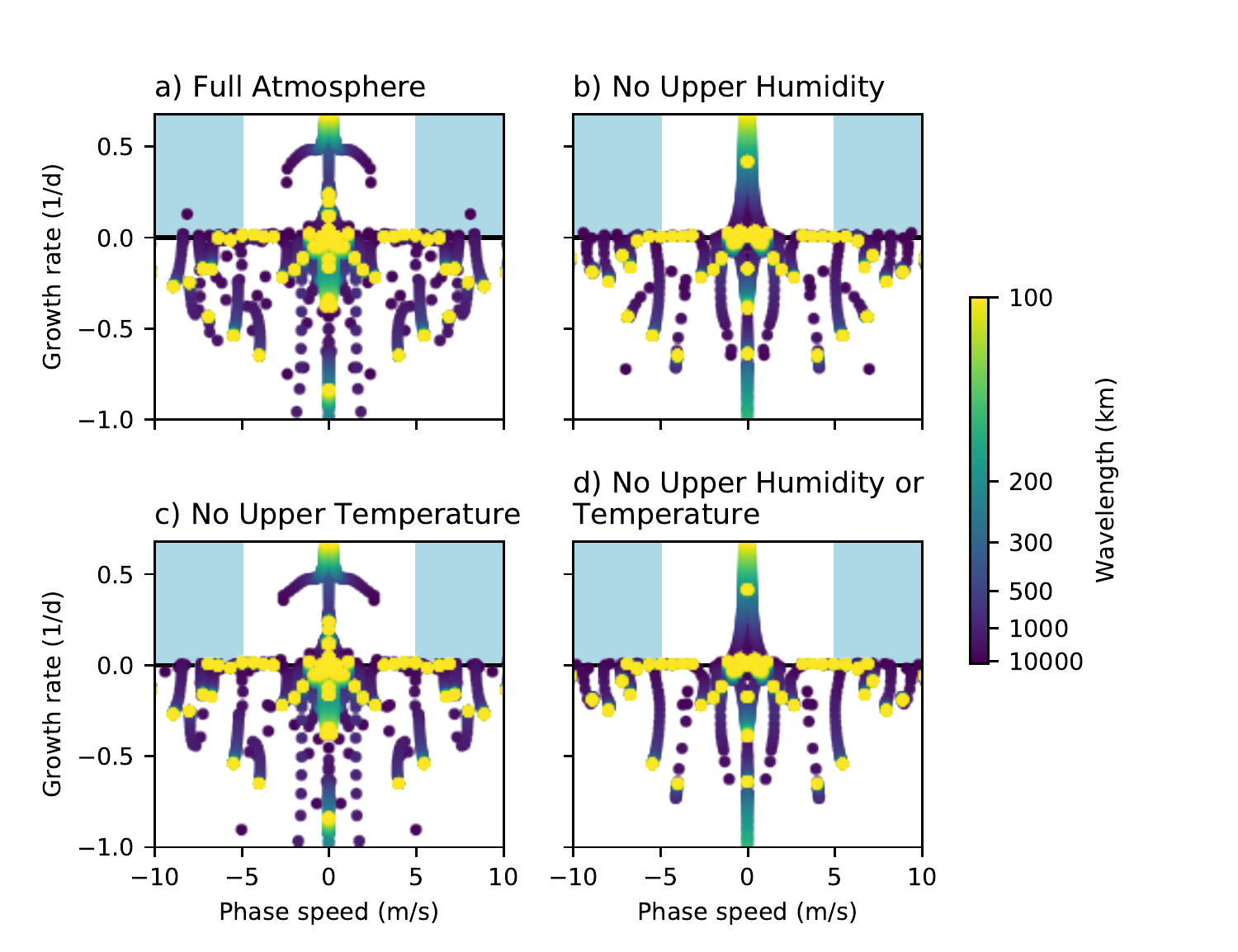}
\caption{Same as Figure \ref{fig:spectra} but only showing modes with a phase
speeds less then 10 m/s.\label{fig:spectra-zoom}}
\end{figure}

Figure \ref{fig:spectra} shows the dispersion
relationships resulting after each of these ablations, a zoomed-in version of which is shown in Figure \ref{fig:spectra-zoom}.
With all atmospheric inputs, there are numerous propagating modes with phase speeds between \SIlist{10;25}{m/s} with positive growth rates .
These modes become increasingly unstable for shorter wavelengths.
When the upper-atmospheric humidity is ablated, there still remain numerous unstable modes, including a ultra-fast \SI{100}{m/s} propagating instability.
The results when the upper-atmospheric temperatures are ablated, but not the upper moisture, are similar to the full atmospheric input.
Finally, ablating both the temperature and humidity inputs from the upper atmosphere removes many unstable propagating modes.
The remaining unstable modes are either stationary or have very slow phase speeds.
This suggests that ablating both humidity and temperature is necessary for nonlinear simulations to be stable when using an ML trained on coarse-grained GCRM data that is at 3-hourly time resolution. Table \ref{tab:info} summarizes these results.

\begin{figure}
\includegraphics[width=1\textwidth]{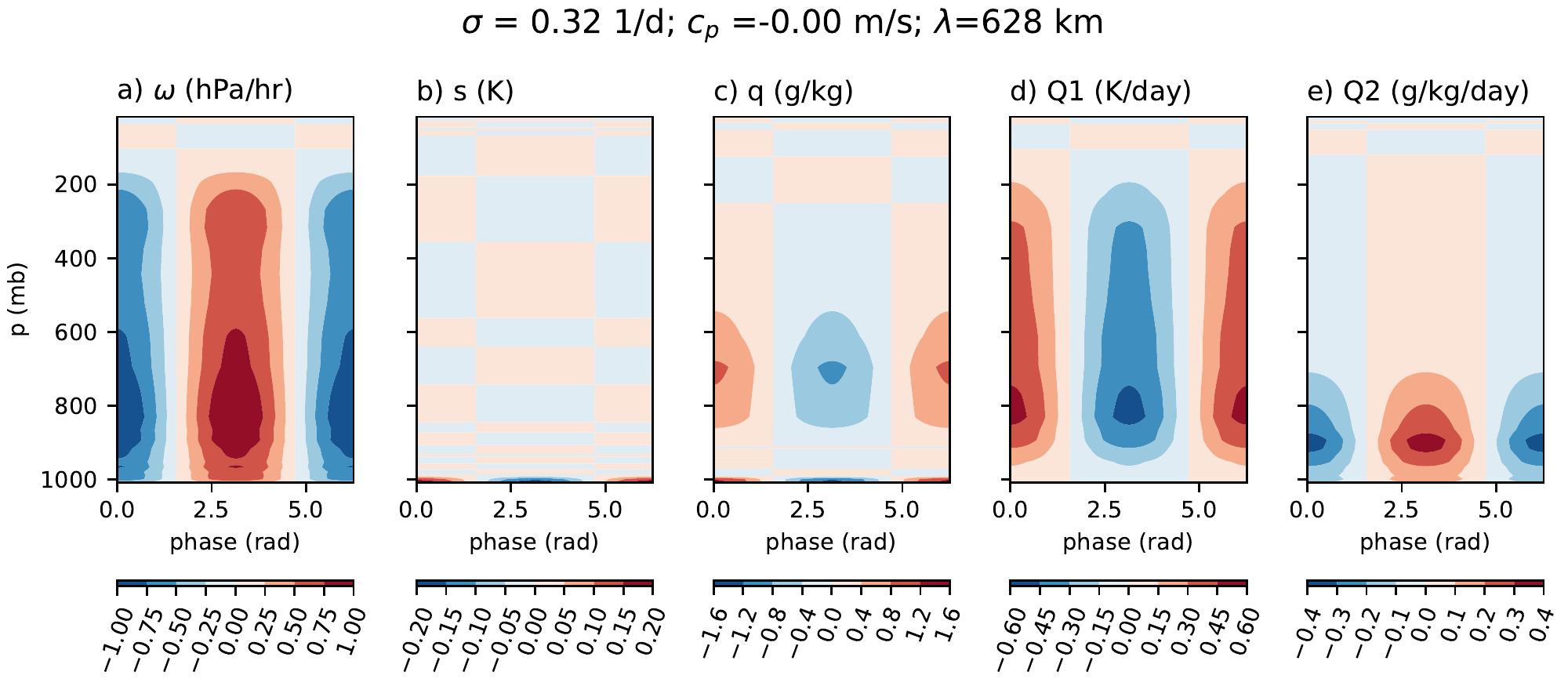}
\caption{Slow-moving instability reminiscent of a moisture mode.\label{fig:standing-instability}}
\end{figure}

\begin{figure}
\includegraphics[width=1\textwidth]{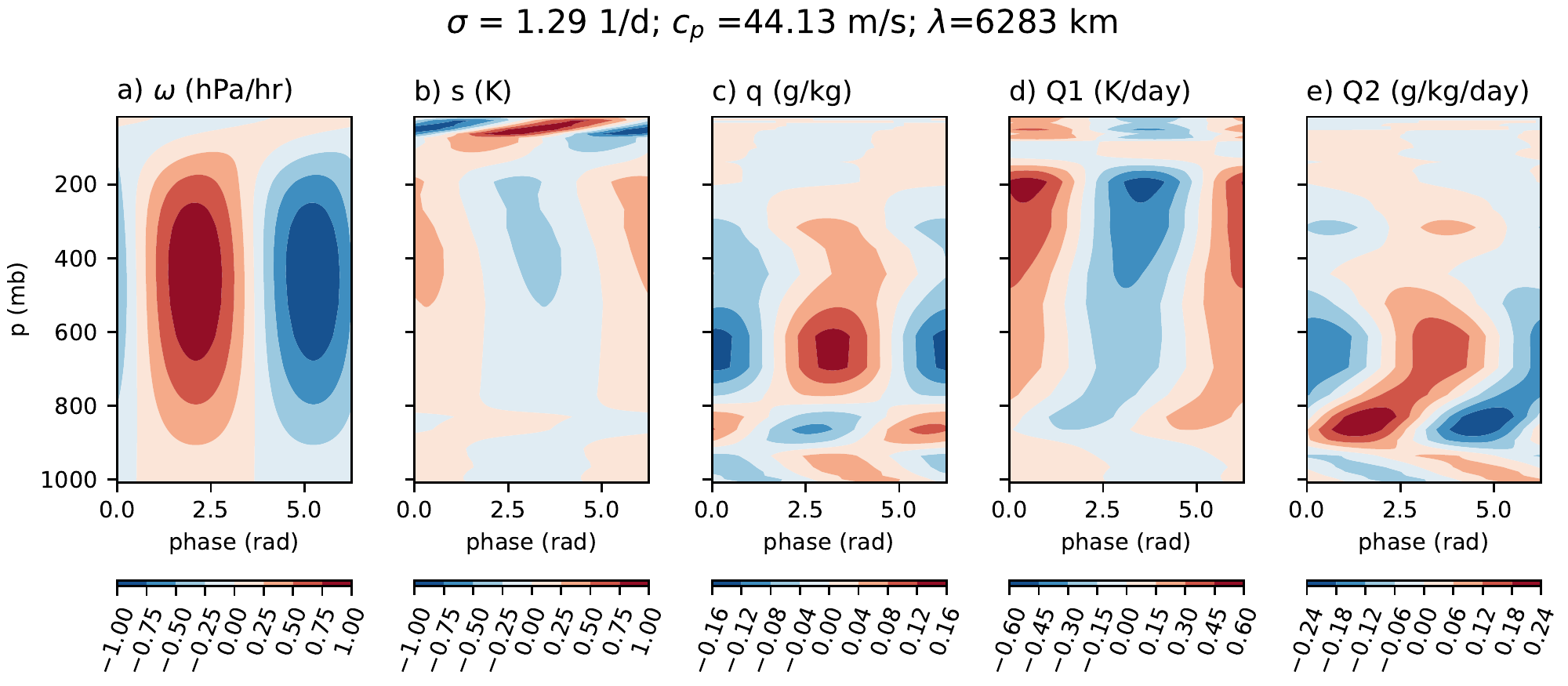}
\caption{Structure of a spurious unstable propagating mode.\label{fig:scary}}
\end{figure}

The phase-space structure of two modes is shown in Figures \ref{fig:standing-instability} and \ref{fig:scary}.
Figure \ref{fig:standing-instability} shows a standing mode of the LRF-wave system in the fully-ablated case.
For a wavelength of \SI{628}{km} this mode has an $e$-folding time of $1/.32 \approx \SI{3}{\day}$ and zero phase-speed.
It primarily couples vertical velocity (panel a) to the total water mixing ratio (panel c), with upward velocity (negative $\omega$) corresponding to anomalous moistening and moisture (panel c).
The heating and cooling (panel d) nearly perfectly compensates for the vertical motions.

The tropospheric heating and temperature are relatively uninvolved, but boundary layer heating and temperature anomalies do have a large magnitude.
Thus, the standing mode appears to be mostly a moisture mode.  Similar modes are responsible for convective self-aggregation in large-domain CRM simulations of radiative-convective equilibrium \citep{Bretherton2005-hr} and are thought to be important for large-scale organized convection such as the Madden Julian Oscillation \citep{Sobel2013-nl,Adames2015-zv}.

\citet{Kuang2018-wh}, found that a similar mode is unstable only when the LRF includes radiative effects. In contrast to Kuang's study, a NN trained to predict $Q_1- Q_{rad}$, where $Q_{rad}$ is the coarse-grained radiative tendency of the \SI{4}{km} model, also predicted an unstable standing mode.
However, it is not clear that this method reliably separates the convection from the radiation because of the noisiness inherent in the GCRM budget residuals and training method.

The LRF-wave analysis can also identify spurious wave-like modes which could contribute to numerical instability in coupled simulations.
Figure \ref{fig:scary} shows a mode of the un-ablated model which has planetary-scale wavelength of \SI{6283}{km}, phase speed of \SI{44}{m/s} and fast growth rate of \SI{1.29}{\per\day}.
This mode is stable for shorter waves, so we have chosen a longer wavelength.
The moisture, humidity, moistening, and drying tendencies are in phase with each other but in quadrature with the $\omega$.
The vertical motion tilts upward and away from the direction of propagation.
Both humidity and temperature anomalies contribute significantly to the wave's structure, but have strange vertical structures.
The upper-atmosphere temperature anomalies are strongly coupled to moisture anomalies in the free troposphere, which have a complex vertical structure.
These structure are not reminiscent of known modes of convective organization, and such a mode could explain the instability this scheme causes when coupled to a GCM.

\begin{table}
\begin{tabular}{>{\centering}p{3cm}>{\centering}p{3cm}>{\centering}p{3cm}>{\centering}p{3cm}>{\centering}p{3cm}}
Upper atmospheric humidity input (above level 15) & Upper atmosphere temperature input (above level 19) & Coupled blow-up & Maximum phase speed  & Standing instabilities\tabularnewline\addlinespace[0.2cm]
\midrule
Yes & Yes & Yes & 50 m/s & yes\tabularnewline
Yes & No & - & 50 m/s  & yes\tabularnewline
No & Yes & - & 100 m/s  & yes\tabularnewline
No & No & No  & 1 m/s  & yes\tabularnewline
\end{tabular}

\caption{Summary of stability issues related to removing upper atmospheric
inputs.\label{tab:info}. The second to last column shows the maximum phase speed over all modes with growth rates larger than \SI{0.05}{\per\day}.}
\end{table}

\subsection{Stabilizing Gravity Wave Modes via Regularization of Inputs\label{sec:res-stable=sp}}

It is natural to wonder whether the wave-coupling framework successfully applied to the GCRM data can also predict prognostic-mode failures in the SPCAM simulation.
The answer is not obvious since ML climate models trained on vastly different data sets exhibit different forms of instability. For instance, the SPCAM-trained "NN-unstable" model tends to go unstable outside of the tropics, and more gradually than the GCRM-trained climate model diagnosed above. This suggests that different ML-based models go unstable for different root causes. Consistent with this view, despite successfully stabilizing the GCRM-trained NN, input ablation does not stabilize the SPCAM-trained ``NN-unstable'': that NN still crashes after $\sim3 $ days when coupled to CAM, even after ablating the input vector's top half components (15 first components of water vapor and temperature profiles, from $p=0\textnormal{hPa} $ to $p=274\textnormal{hPa} $, see e.g. Movie S2). 
As an alternative to ablation, we introduce a new empirical technique, ``input regularization'', that can improve the SPCAM-trained NN's prognostic performance.
We then show that the computationally affordable wave-coupling introduced in Section \ref{sec:lrf+wave+intro} can predict how much regularization is required to stabilize coupled GCM-NN simulations.

We begin by revisiting the physical credibility of NN-unstable compared to NN-stable using the diagnostics that previously revealed the causal ambiguity endemic to the GCRM-trained NN. From this view, we expect NN-unstable to struggle in prognostic mode, since it exhibits significant positive heating and moistening biases (see Figure S1) for large mid-tropospheric moisture (>20mm) and low LTS (<12K), conditions favorable for deep convection. Positive convective moistening biases in moist regions may then lead to instability through spurious moistening of growing water vapor perturbations. 

\begin{figure}
\includegraphics[width=1\textwidth]{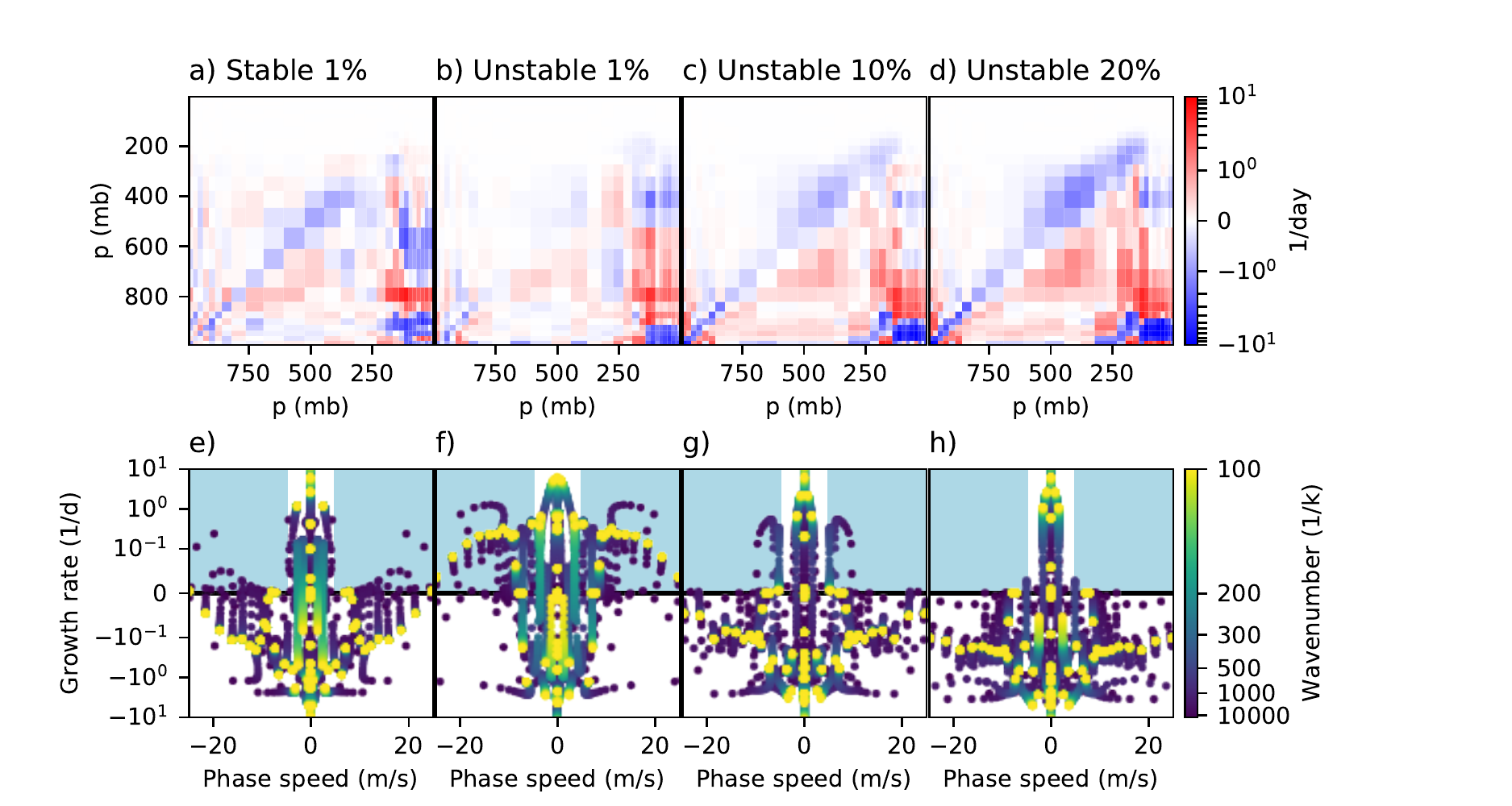}
\caption{(Top) Regularized $\left(\partial\boldsymbol{\dot{q}_{v}}/\partial\boldsymbol{q_{v}}\right) $ linear response functions (in units 1/day) of NN-stable (leftmost column) and NN-unstable (three rightmost columns) for various regularization amplitudes (in $\% $). (Bottom) Corresponding stability diagrams obtained by coupling the linear response functions to simple two-dimensional dynamics. As in Figures \ref{fig:spectra} and \ref{fig:spectra-zoom} the light-blue background indicates where the phase speed is greater than $|c_p| > \SI{5}{m/s}$ and the growth rate is positive.\label{Regularized_LRF}}
\end{figure}

Next, we study the effect of increased input regularization as described in Sec \ref{sec:regularization}, focusing on the relation between offline prediction
and online performance.
Figure \ref{Regularized_LRF} shows that input regularization effectively
controls stability properties.
The top row shows the regularized LRFs calculated about the base state
$\boldsymbol{x_{\mathrm{unstable}}} $ for regularized amplitudes of $1\% $,
$10\% $, and $20\% $. 
A preliminary stability analysis, including direct eigenvalue
analysis and simple dynamical coupling using the strict weak-temperature
gradient approximation \citep{Beucler2018-cd}, indicates that
NN-unstable's damping rates are closer to 0 than NN-stable's damping rates. 
Although these results suggest that NN-unstable is unable to damp developing
instabilities quickly enough, the stability diagram from our wave-coupler
(Figure \ref{Regularized_LRF}f) can provide a more extensive
description of developing instabilities if we use an input ensemble tightly
regularized ($1\% $) about the base state $\unstable$. 
While NN-stable only has a few slowly-growing modes about the unstable base
state (Figure \ref{Regularized_LRF}e), NN-unstable exhibits a myriad of fast-growing ($\sim1\mathrm{day^{-1}} $)
modes  (Figure \ref{Regularized_LRF}f) propagating at phase speeds between 5m/s and 20m/s, indicated with the light-blue background. 
Hence our wave-coupling framework can anticipate the instability arising from coupling
NN-unstable to CAM.

\begin{figure}
\includegraphics{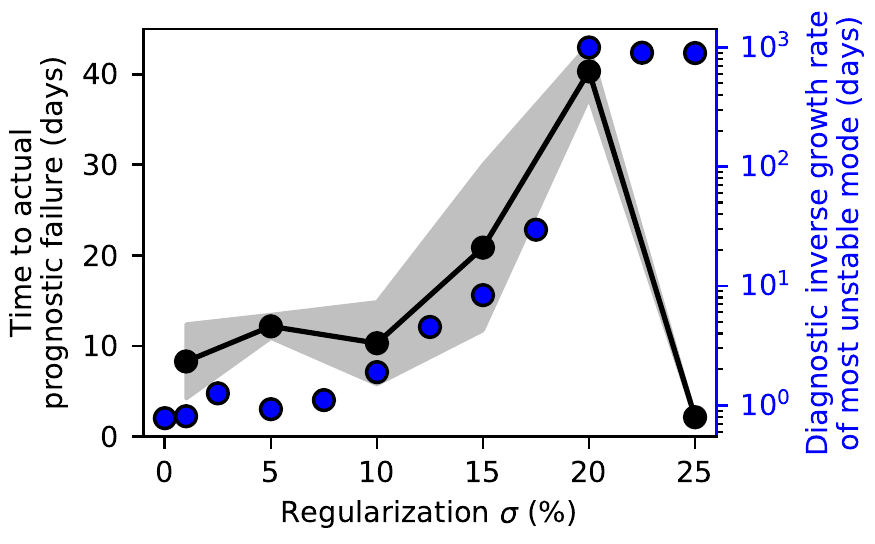}
\caption{Time to failure for an ensemble of prognostic tests (ensemble mean in black and standard deviation in grey) with varying ``input regularization'' spread magnitude (see text) compared to (blue) offline predictions from the most unstable mode derived from the 2D wave-coupler, defined as the maximal growth rate from the stability diagram that propagates with a phase speed of absolute value greater than 5 m/s. \label{fig:NNCAM_prognostic_summary}}
\end{figure}

Interestingly, our wave-coupling framework further predicts that more input regularization should stabilize NN-unstable.  The largest propagating growth rates decrease from $ \sim1\;\mathrm{day^{-1}}$ for a $1\% $~regularization amplitude to $\sim10^{-3}\;\mathrm{day^{-1}} $ for a $25\% $~regularization amplitude  (Figure \ref{Regularized_LRF}h). 
To test this prediction, we run a suite of CAM simulations in which the host climate model's grid columns are each coupled to the mean prediction of a 128-member ensemble of NN predictions, formed via Gaussian-perturbed inputs, instead of the typical single NN prediction per grid cell. The amount of input spread is varied between 1\% and 25\% standard deviation across five experiments. To provide a measure of internal variability, each experiment is repeated across a four-member mini-ensemble launched from different SPCAM-generated initial conditions spaced five days apart. 
Figure \ref{fig:NNCAM_prognostic_summary} shows the time to failure of these
runs for increasing regularization.
With 15$\% $ or less input regularization, none of these simulations is able to
run more than 21 days, consistent with the existence of many unstable modes in
the 2D wave-coupler diagnostic; instead, variants of the same extratropical
failure mode eventually occur. 
But when 20\% spread is used to seed sufficient diversity in the input
conditions, longer term prognostic tests become possible.
Interestingly, the most dramatic effect of the input regularization in delaying
the time to instability happens between spread magnitudes between 15\% and
20\% -- this is consistent with the fact that the offline 2D wave coupler
tests indicate an especially prominent shutdown of unstable modes in the
vicinity of a 16\%  regularization magnitude. 
While we do not expect a simple linear model neglecting rotation to accurately
predict non-linear, mid-latitude instabilities of a full-physics general
circulation model, our results suggest that the offline diagnostics tools
developed in this study apply to a wide range of instability situations.

\section{Conclusions\label{sec:conclusion}}

Machine learning parameterizations promise to resolve many of the structural biases present in current traditional parameterizations by greatly increasing the number of tuning parameters.
This massive increase in flexibility has two main drawbacks: 1) ML parameterizations are not interpretable a priori and 2) neural network parameterizations are often unstable when coupled to atmospheric fluid dynamics. This study addresses both of these points by developing an interpretability framework specialized for ML parameterizations of convection, and deepening analysis of the relationship between offline skill vs. online coupled prognostic performance.

By systematically varying input variables in a two-dimensional thermodynamic space, we have demonstrated that two independent ML parameterizations behave as our intuition would expect. Increases in lower-tropospheric moisture tend to greatly increase the predicting heating and moistening, a widely documented fact \citep{Bretherton2004-jf}, while
increasing the lower-tropospheric stability effectively controls the depth of convection.
These changes are consistent with the actual sensitivity present in our training dataset, and both the SPCAM and GCRM neural networks behave somewhat consistently, demonstrating the robustness of the machine learning approach to parameterizations.
They both predict that net precipitation increases for moist and unstable columns, although the precise vertical structures of the their heating and moistening profiles differ. 
Future work in a similar spirit could easily build on these methods. For instance, a limitation of our approach here is that lower-tropospheric stability and moisture co-vary strongly because stable columns tend to be warmer and therefore carry more moisture. Therefore, the sensitivities we demonstrate are not entirely independent.
Instead of stability and moisture, the estimated inversion strength \citep{Wood2006-ro} and lower-tropospheric buoyancy \citep{Ahmed2018-yd} may control convection more orthogonally, which would further ease the interpretation of such results, and is recommended in future diagnostics of this vein.

We have also developed an offline technique that shows some skill in predicting whether a given ML parameterization will be stable online (i.e. when coupled to atmospheric fluid dynamics) in both the GCRM-trained and SPCAM-trained limits, despite their many intrinsic differences. By coupling gravity-wave dynamics to the linearized response functions of an ML parameterization, one can compute the growth rates, phase speeds, and physical structure of the gravity wave modes associated with the parameterizations.
We find that, in both SPCAM and SAM, propagating unstable modes are associated with the numerical instability in online coupled runs, and likely one root cause of instability. In stabilized versions of both schemes (using ``input ablation'' for SAM and ``input regularization'' for SPCAM), the propagating modes are all stable and prognostic tests run more stably.
That this framework does not include rotation or background shear indicates that interactions between gravity waves and the parametrized heating play a role in numerical instability. We speculate such instability causes coupled simulations to drift towards the boundary of their training datasets. Once they reach this boundary, the neural networks are forced to extrapolate to unseen input data, which causes the final rapid blow-up.

Since this is also the first study to comprehensively compare NNs trained on coarse-grained GCRM data vs. SPCAM data, some comments on interesting differences worthy of future analysis are appropriate. Fully understanding why the SPCAM LRFs are so much noisier than the SAM LRFs will require further study of the many factors that could be involved, beyond obvious differences in the nature of the training data (e.g. hyperparameter settings and neural network architecture and optimization settings), especially since computing LRFs from traditional parameterizations typically requires some degree of regularization \citep[e.g.,][]{Beucler2019-lv,Kuang2007-ph}. Multiple techniques have been developed to smooth neural-network Jacobians, including averaging Jacobians derived from an ensemble of neural networks \citep{krasnopolsky2007reducing}, taking the Jacobian of a single mean profile \citep{chevallier2001evaluation}, or even multiplying the Jacobian by a ``weight smoothing'' regularization matrix \citep{aires1999weight}. Here, we have introduced ``input regularization'' as an alternative strategy to ensemble-average Jacobians without having to train multiple networks. The regularized SPCAM LRFs are smoother, making them attractive for physical interpretation and easier to compare to GCRM LRFs, as well as easier to analyze using our wave-coupling framework. But we caution that, while incrementally helpful for full prognostic stability, ``input regularization'' should not be viewed as a solution to the instability problems in SPCAM-trained models. More attention on other strategies like formal hyperparameter tuning to efficiently uncover optimally skillful fits, which may be even more likely to perform stably online, is also warranted.

This wave-coupling analysis also has potentially interesting physical implications, but more work is required to compare and contrast NN-derived LRFs with other approaches \citep[e.g.]{Kuang2018-wh}.
Unlike \citet{Kuang2018-wh}, our analysis is not conducted about radiative-convective equilibrium (RCE) profiles, but rather about base states close to the regions where we saw numerical instabilities.
For a fair comparison, we should compute our spectra about an equilibrium state, if such exists for our schemes. Finally, the robustness of our spectra is hard to quantify, especially for phase-speeds and growth rates near zero. This is acceptable for discovering the spurious unstable propagating waves above, but making inferences about true physical modes will require a more rigorous statistical framework.

In summary, this manuscript has presented a pair of techniques that allow peering into the inner workings of two sets machine learning parameterizations. These tools have led to the development of new regularization techniques, and could allow domain experts to assess the physical plausibility of an ML parameterization. Reassuringly, ML parameterizations appear to behave according to our physical intuition, creating the potential to accelerate current parameterizations and develop more accurate data-driven parameterizations. 








%

%
\paragraph*{Acknowledgments}
When starting this work, N.B. was supported as a postdoctoral fellow by the Washington Research Foundation, and by a Data Science Environments project award from the Gordon and Betty Moore Foundation (Award \#2013-10-29) and the Alfred P. Sloan Foundation (Award \#3835) to the University of Washington eScience Institute. C.B. was initially supported by U. S. Department of Energy grant DE-SC0016433. NB and CB acknowledge support from Vulcan Inc. for completing this work. TB and MP acknowledge support from NSF grants OAC-1835863, OAC-1835769, and AGS-1734164, as well as the Extreme Science and Engineering Discovery Environment supported by NSF grant number ACI-1548562 (charge numbers TG-ATM190002 and TG-ATM170029) for computational resources.

%
\appendix{}
\section{Derivation of 2D anelastic wave dynamics}
\label{sec:appendix}

\subsection{Continuous equations}

The linearized hydrostatic anelastic equations in the horizontal direction $x$ and height $z$ are given by 

\begin{align*}
q_{t}+\bar{q}_{z}w=Q_{2}',\\
s_{t}+\bar{s}_{z}w=Q_{1}',\\
u_{t}+\phi_{x}=-du.\label{eq:progs}
\end{align*}
The prognostic variables are humidity $q$, dry static energy $s=T + \frac{g}{c_p} z$, horizontal velocity $u$, and vertical velocity $w$. These are assumed to be perturbations from a large scale state denoted by $\bar{\cdot}$. 
The an-elastic geopotential term is given by $\phi=p'/\rho_0$, where $\rho_0(z)$ is a reference density profile specified for the full nonlinear model.

These prognostic equations are completed by assuming hydrostatic balance and mass-conservation.
Hydrostatic balance is given by 
\[ \phi_z = B \]
where the $B=gT/\bar{T}$ is the buouyancy.
Mass conservation is defined by
\[ u_x + \frac{1}{\rho_0}\partial_z \rho_0 w .\]

We now combine these diagnostic relations and zonal momentum equation into a single prognostic equation for $w$.
For convenience, we define two differentiable operators, $L=\partial_z$ and ${\cal H}=\frac{1}{\rho_0}\partial_z \rho_0$. Taking the $x$ derivative of the momentum equation, and applying the divergence-free condition gives 
\[
{\cal H}w_{t}+d{\cal H}w-\phi_{xx}=0.
\]
Then, applying $L$ gives 
\[
L{\cal H}(\partial_t + d)w = B_{xx}.
\]
We let $A=L{\cal H}$, and manipulate the equations to obtain 
\[ 
w_t = -\partial_{xx} A^{-1} B -d w.
\]
Because $A$ is an elliptic operator in the vertical direction, it requires two boundary conditions. In this case, we assume these are given by a rigid lid and impenetrable surface (e.g. $w(0)=w(H_T)=0$) where $H_T$ is the depth of the atmosphere.

\subsection{Vertical Discretization}
\label{sec:discrete-elliptic}

Solving (\ref{eq:w}) numerically requires discretizing the elliptic operator $A$. To do this, we assume that  $w$, $s$, and $q$ are vertically collocated. Then, in the interior of the domain, the operator $A$ can be discretized as the following tri-diagonal matrix:
\[
(Aw)_{k}=a_{k} w_{k-1}+b_{k}w_{k}+c_{k}w_{k+1}
\]
where
\begin{align*}
    a_{k}&=\frac{\rho_{k-1}}{(z_{k}-z_{k-1})(z_{k+1/2}-z_{k-1/2})\rho_{k-1/2}}\\ b_{k}&=-\frac{\rho_k}{(z_{k+1/2}-z_{k-1/2})}\left[\frac{1}{(z_{k+1}-z_{k})\rho_{k+1/2}}+\frac{1}{(z_{k}-z_{k-1})\rho_{k-1/2}}\right],\\ c_{k}&=\frac{\rho_{k+1}}{(z_{k+1}-z_{k})(z_{k+1/2}-z_{k-1/2})\rho_{k+1/2}}.
\end{align*}
The index $k$ ranges from 1 to $N$, the number of vertical grid cells, and $z$ is the height.

The rigid-lid boundary conditions are satisfied by: $w_0 = - w_1
$ and $w_{n+1}=-w_{n}$. It is not simply $w_{0}$ because the vertical
velocity should be located at the cell center. These boundary conditions can be implemented by modifying the matrix representation of $A$ to satisfy
\begin{align*}
    (Aw)_{1}=-a_{1}w_{1}+b_{1} w_{1}+c_{1}w_{2},\\
    (Aw)_{n}=a_{n}w_{n-1}+b_{n}w_{n}-c_{n}w_{n}
\end{align*}
at the lower and upper boundaries.


\end{document}